\documentstyle[epsf,twocolumn,aps]{revtex}
\tightenlines

\newcommand{\be}{\begin{equation}}
\newcommand{\ee}{\end{equation}}

\newcommand{\mgbb}{MgB$_2$}

\begin{document}
\twocolumn[\hsize\textwidth\columnwidth\hsize\csname @twocolumnfalse\endcsname


\title{Superconductivity of \mgbb: Covalent Bonds Driven Metallic}

\author{J. M. An and W. E. Pickett} 

\address{Department of Physics, University of 
                     California, Davis CA 95616}

\date{\today}
\maketitle
\tightenlines
\begin{abstract}
A series of calculations on \mgbb~and related isoelectronic 
systems indicates that 
the layer of Mg$^{2+}$ ions
lowers the non-bonding B $\pi$ ($p_z$) bands relative to the bonding
$\sigma$ ($sp_xp_y$) bands compared to graphite, causing $\sigma \rightarrow
\pi$ charge transfer and $\sigma$ band doping of 0.13 holes/cell. 
Due to their 
two dimensionality the $\sigma$ bands contribute 
strongly to the Fermi level density of states.
Calculated deformation potentials of $\Gamma$ point phonons identify
the B bond stretching modes as dominating the electron-phonon coupling.
Superconductivity driven by $\sigma$ band holes is
consistent with the report of 
destruction of superconductivity by doping with Al.

\end{abstract}
\vskip 1cm
\footnoterule
\newpage

]
The dependence of the superconducting critical temperature T$_c$ on
structure is unclear after decades of study, and recent discoveries 
have further confused the issues.  In conventional superconductors
(the high T$_c$ cuprates comprise a special case) it
has been generally thought that high symmetry, preferably cubic, is
favorable for higher T$_c$.  This trend has held up in elemental
superconductors (Nb, La under pressure, at 9 K and 13 K, 
respectively),
for binaries (Nb$_3$(Al,Ge), 23 K\cite{gavaler}), (pseudo)ternaries such as 
(Ba,K)BiO$_3$,\cite{bkbo} and even fullerene superconductors 
(T$_c$ to 40 K\cite{cs3c60}).

Recently several intriguing counterexamples to this trend have come
to light.  Semiconducting HfNCl (and ZrNCl) is a van der Walls bonded set of 
covalent/ionic bonded layers, but superconducts up to 25 K when doped
(intercalated) with alkali metals.\cite{MNCl}  There is evidence that 
the surface layer of solid C$_{60}$
becomes superconducting up to 52 K when it is injected to high hole 
concentrations.\cite{c60}  And most recently, it is
reported that the layered metal/metalloid compound \mgbb~superconducts
at $\sim$40 K,\cite{akimitsu,budko} 
which is by far the highest T$_c$ for a binary system.  
The B isotope shift of T$_c$ reported by Bud'ko {\it et al.}\cite{budko}
and most other early experimental data\cite{other}
suggests conventional BCS strong-coupling s-wave electron-phonon (EP) pairing.
These examples suggest there are important aspects of two dimensionality
(2D) for conventional superconductors that are yet to be understood.

A close analogy for \mgbb, structurally, electronically,  
and regarding superconductivity,
is graphite.  Graphite has the same C layer structure as B has in \mgbb.
The layer stacking that occurs in graphite is central to its semimetallic 
character but is not important in our analogy.
Graphite is isoelectronic with \mgbb~; previous studies have established
that the Mg atom is effectively ionized.
Finally, graphite becomes superconducting up to
5 K, but only when doped (intercalated).\cite{c2}  
Both graphite and \mgbb~have planar $sp^2$ bonding, and in graphite it
is well established that
three of carbon's four valence electrons are tied up in strong $\sigma$ bonds
lying in the graphite plane, and the other electron lies in 
non-bonding ($p_z$)
$\pi$ states.  Electron doping of graphite, 
achieved by intercalating
alkali atoms between the layers (Na is the most straightforward case) leads
to occupation of otherwise unfilled $\pi$ states and leads to T$_c$ as
high as 5 K.  \mgbb, on the other hand, superconducts at $\sim$40 K 
at stoichiometry.

The light masses in \mgbb~enhance the
phonon frequency ($\omega_{ph} \propto M^{-1/2}$) that sets the 
temperature scale of T$_c$ in BCS theory.  Even considering this tendency,
there must be some {\it specific feature(s)} that produces such a remarkable
T$_c$, and moreover does so with no $d$ electrons, nor even the 
benefit of a density of states (DOS) peak.\cite{kortus,satta} 
In this paper we identify
these features: \\
(1) hole doping of the covalent $\sigma$ bands, achieved through the ionic,
layered character of MgB$_2$,\\
(2) 2D character of the $\sigma$ band density of states, making 
small doping concentrations $n_h$ give large effects 
N($\varepsilon_F$)$\sim m^*/\pi \hbar^2$ independent of $n_h$, and \\
(3) an ultrastrong deformation potential of the $\sigma$ bands from the
bond stretching modes. \\
In this paper we exploit the
similarities between graphite and \mgbb~to build a basic understanding of
its electronic structure, 
then focus on the differences that are connected to strong
EP coupling, and finally provide an estimate of T$_c$ that indicates the
picture we build can account for the observations.
This picture further predicts that electron doping by 
$\sim$0.2 carriers per cell
will strongly affect T$_c$ adversely, a result that has been reported
by Slusky {\it et al.}\cite{cava}
Although Hirsch has also focussed on the hole character of
the $\sigma$ bands,\cite{hirsch} his emphasis is otherwise quite
different from that described here.  

Calculations of the electronic structure have been done using
the linearized augmented plane wave (LAPW) method\cite{djsbook}
that utilizes a fully general shape of density and potential,
as implemented in the WIEN97 code.\cite{wien}
Experimental lattice constants of a=3.083~\AA, c=3.521~\AA~were used.
LAPW sphere radii (R) of 2.00 a.u. and 1.65 a.u. were chosen for
the Mg and B atoms, respectively,
with cutoff RK$_{max}$=8.0, providing
basis sets with more than
1350 functions per primitive cell.
The generalized gradient approximation 
exchange-correlation functional of Perdew {\it et al.}
\cite{gga} was used in the present work.

\vskip -5mm
\begin{figure}[tbp]
\epsfxsize=7.0cm\centerline{\epsffile{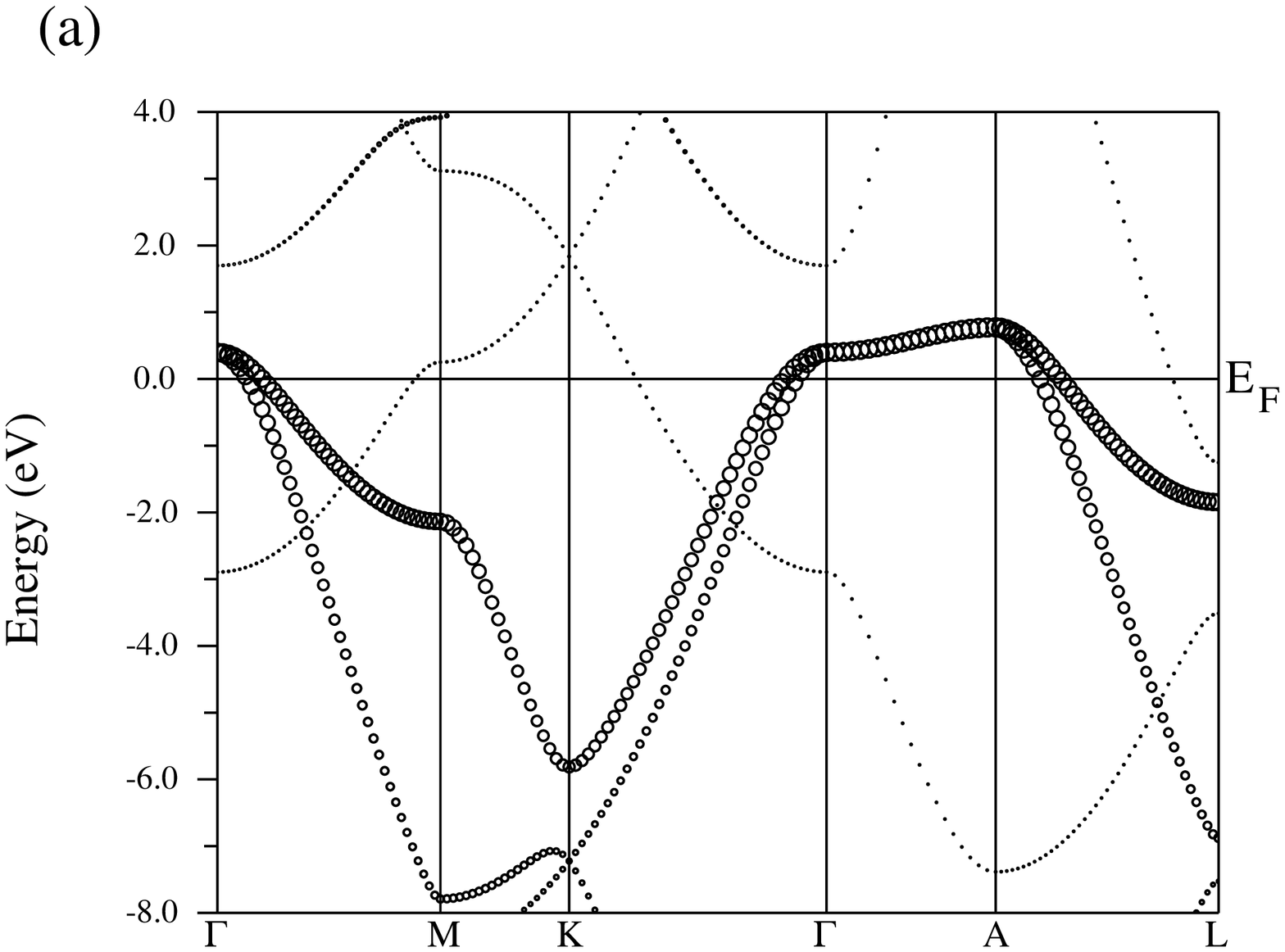}}
\vskip -17mm
\epsfxsize=7.0cm\centerline{\epsffile{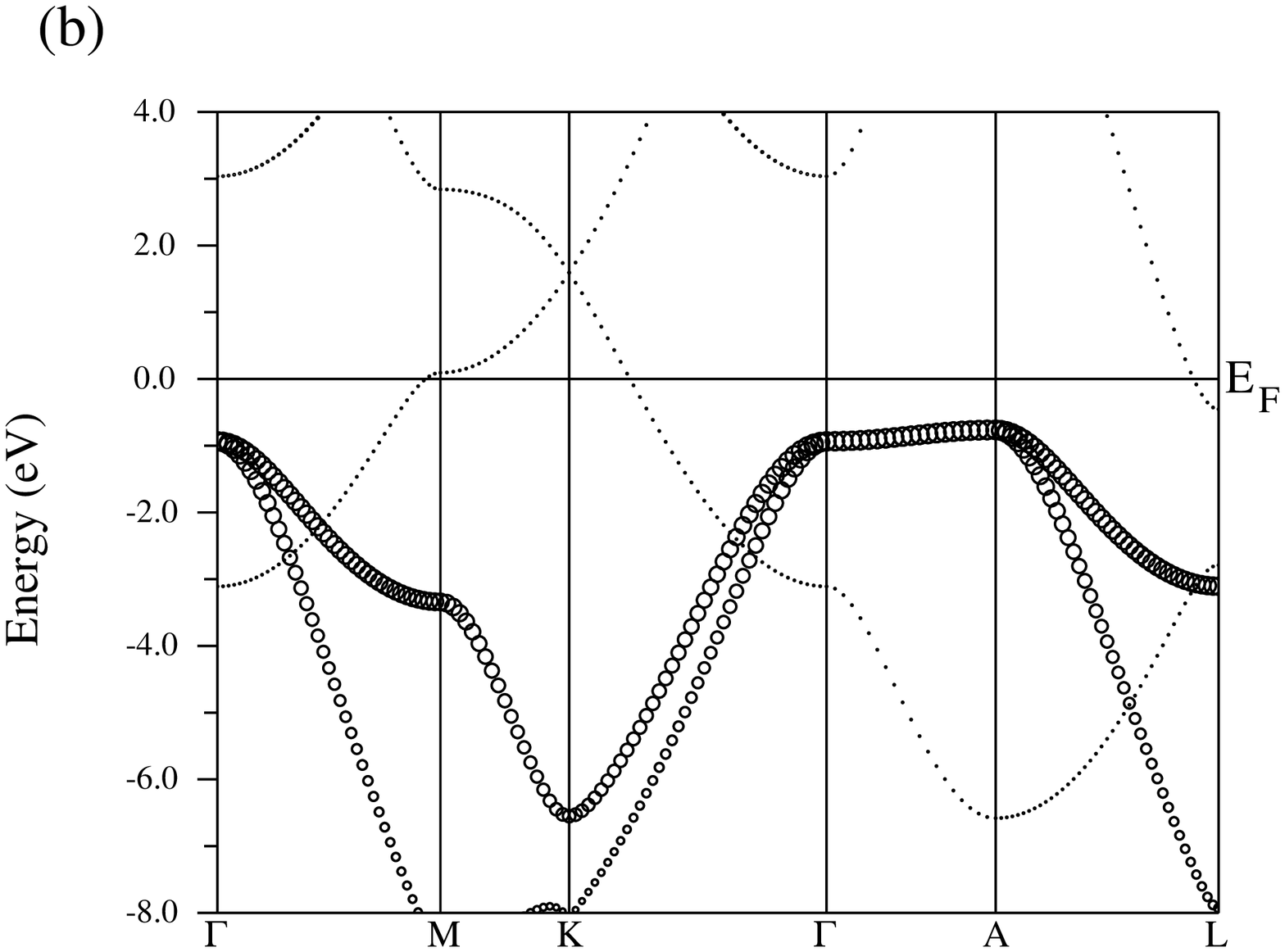}}
\vskip -17mm
\epsfxsize=7.0cm\centerline{\epsffile{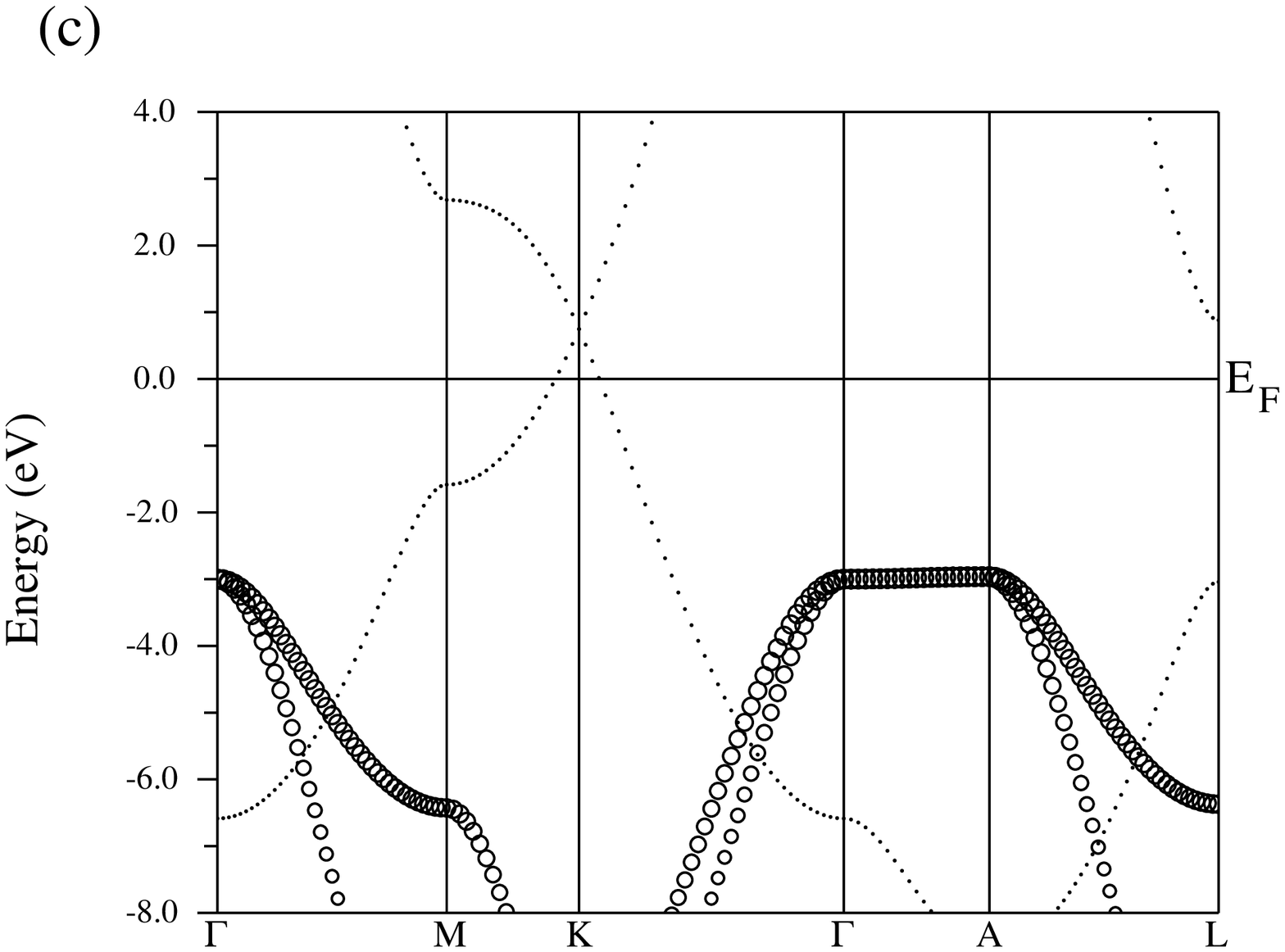}}
\vskip -8mm
\caption{Band structure along main hexagonal symmetry lines, for
(top) \mgbb, (middle) $\Box^{2+}$B$_2$, (bottom) primitive graphite C$_2$.
The planar $\sigma$ states, highlighted with larger symbols, fall
in energy in this progression, and only in \mgbb~are they partially
unoccupied.  The point A = (0,0,$\pi/c$) is perpendicular to the
($k_x,k_y$) plane.
}
\label{Bands1}
\end{figure}

The band structure of \mgbb~is shown in Fig. \ref{Bands1} 
(top panel) in comparison
with that of primitive graphite (bottom panel) with a single 
layer per cell like the B$_2$
sublattice in \mgbb.  For each two distinct sets of bands are 
identifiable: the highlighted
$sp^2~(\sigma)$ states, and the $p_z~(\pi)$ states.
The striking difference is in the position of 
the $\sigma$ bands, which is evident in Fig. \ref{Bands1}.
Whereas the 
$\sigma$ bonding states are completely filled in graphite and
provide the strong covalent bonding, in \mgbb~they
are unfilled and hence metallic, 
with a concentration of 0.067 holes/B atom
in two fluted cylinders surrounding the $\Gamma$-A line of the Brillouin 
zone.\cite{kortus}
There are correspondingly more electron carriers in the 
$\pi$ bands.  This
decrease in occupation on the strongly bonding $\sigma$ bands partially 
accounts for the greatly increased planar 
lattice constant of \mgbb~(3.08 \AA) compared to graphite (2.46 \AA).
Our results agree with previous conclusions that MgB$_2$ can be well
characterized by the ionic form Mg$^{2+}$(B$_2$)$^{2-}$.

To identify the origin of the relative shift of the $\sigma$ and $\pi$
bands by $\sim$3.5 eV between graphite and \mgbb, we have considered a
fictitious system $\Box^{2+}$B$_2$ in which the Mg ion is removed but
the two electrons it contributes are left behind (and compensated by a
uniform background charge).  The band structure, shown in the
middle panel of Fig. \ref{Bands1}, is very
similar, except the energy shift of $\sim$1.5 eV downward with respect
to \mgbb~completely fills the
$\sigma$ bands, as in graphite.  This shift is the result of the lack of
the attractive Mg$^{2+}$
potential in \mgbb, which is felt more strongly by the $\pi$ electrons than by
the in-plane $\sigma$ electrons: the attractive potential of Mg$^{2+}$
between B$_2$ layers lowers the $\pi$ bands, resulting in 
$\sigma \rightarrow \pi$ charge transfer that drives the hole doping 
of the $\sigma$ bands.  Belashchenko {\it et al.}\cite{bela} have also
considered a sequence of materials to come to related conclusions about
the band structure, but they did not use isoelectronic systems as
has been done here.

The $\sigma$ bands are strongly 2D (there is very little dispersion along
$\Gamma$-A), but it will be important to
establish the magnitude and effects of interplanar coupling.
The light hole and heavy hole $\sigma$ bands in \mgbb~can be modeled 
realistically in the region of interest (near and above $\varepsilon_F$)
with dispersion of the
form
\be
\varepsilon_k = \varepsilon_{\circ} - \frac{k_x^2 + k_y^2}{2m^{*}}
  - 2 t_{\perp} cos(k_z c),
\ee
where the planar effective mass $m^{*}$ is taken to be positive and $t_{\perp}$
= 92 meV is the small dispersion perpendicular to the layers.
The light and heavy hole masses are 
$m_{lh}^*/m$=0.20, $m_{hh}^*/m$= 0.53, and the mean
band edge is $\varepsilon_{\circ}$=0.6 eV.  
In general, the in-plane ($v_{xy}$) and perpendicular ($v_z$) Fermi velocities 
are expected to be anisotropic: $v_{xy} \sim k_F/m^*$,
$v_z \sim 2 c t_{\perp}$ where $t_{\perp}$ is small.  Near the band edge
($k_F \leq 2 m^* c t_{\perp}$) this anisotropy becomes small, and this is 
roughly the case in MgB$_2$.
The $\pi$ bands are also
effectively isotropic.\cite{kortus,satta}

\begin{figure}[t]
\epsfxsize=6.5cm\centerline{\epsffile{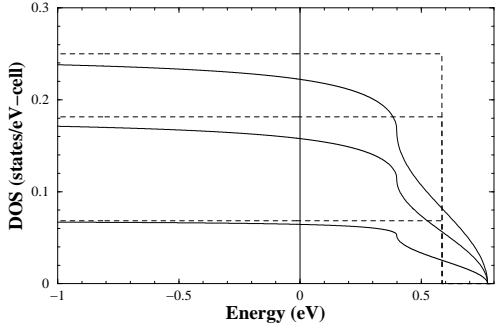}}
\caption{Density of states (solid lines) of the 
light hole band, the heavy hole
band, and the total, for the model of Eq. (1). 
Note that N($\varepsilon$) drops rapidly only 0.4 eV
above $\varepsilon_F$. The dashed lines give the 2D
($t_{\perp}$=0) analogs.   
}
\label{DOS}
\end{figure}

Now we discuss why 
the quasi-2D character of the $\sigma$ bands is an important feature 
of MgB$_2$ and its superconductivity.
Neglecting the $k_z$ dispersion, the 2D hole density of states is 
constant: N$_h^{\circ}$($\varepsilon$)
=$\frac{m_{lh}^* + m_{hh}^*}{\pi \hbar^2}$= 0.25 
states/eV-cell, {\it independently of the fact that the hole doping level
is small}.  The $k_z$ dispersion has only the small effect displayed in Fig.
\ref{DOS}, where the discontinuity in the quasi-2D DOS is seen to be
broadened by $\sim 2 t_{\perp}$.  For MgB$_2$
the $\sigma$ band contribution to N($\varepsilon_F$) is reduced by
about 10\% by $k_z$ dispersion.

If superconductivity is primarily due to the existence of holes in the
$\sigma$ band, and we provide evidence for such a picture below, 
then the DOS in Fig. \ref{DOS} suggests that electron doping
will decrease N($\varepsilon_F$). The decrease will be 
smooth to a doping level
corresponding
to an increase by 0.4 eV of the Fermi level.  i
Then N($\varepsilon_F$) should drop
precipitously with further doping.  
A rigid band estimate gives a value of $n_{h,cr}$
=0.08 electrons necessary to fill
the $\sigma$ bands to the crossover region.  Electrons must also be added
to the $\pi$ bands that lie in the same temperature range, making the total
doping level $n_{cr} \approx$ 0.25 per cell.  

Fig. \ref{DOS} therefore suggests a critical region of electron
doping around 0.25 carriers per cell.
Slusky {\it et al.} have reported\cite{cava} a study of such electron 
doping in the
Mg$_{1-x}$Al$_x$B$_2$ system.
T$_c$ drops smoothly up to $x$=0.1, beyond
which point a two phase mixture of B-rich and Al-rich phases occurs.  At
$x$ = 0.25 and beyond, a single phase, non-superconducting system is
restored, strongly supporting our picture that filling the $\sigma$ bands
will destroy superconductivity.  Using the rigid
band picture, filling them
decreases T$_c$ moderately initially, but as the $\sigma$ bands
become nearly  filled, the coupling decreases abruptly and T$_c$ vanishes, as
observed.  Although our results do not bear directly on the two-phase
question, we note that this occurs just as the $\sigma$ bands are
filling.  At this point N($\varepsilon_F$) is dropping, which normally
favors stability.  The observed instability suggests that a very
small density of $\sigma$ holes that are very strongly coupled to the
lattice is what underlies the lattice instability.

Now we address the question of EP coupling strength $\lambda$.
The calculated value of N($\varepsilon_F$)
=0.71 states/eV-cell correponds to a bare specific heat coefficient
$\gamma_{\circ}$ = 1.7 mJ/mole-K$^2$.  The available experimental estimate
is\cite{budko} $\gamma_{exp} = (1+\lambda) \gamma_{\circ}$ = 3 $\pm$ 1
mJ/mole-K$^2$, giving $\lambda_{exp} \sim 0.75$.
Kortus {\it et al.} have used the rigid muffin tin
approximation\cite{GG} to
obtain an idea of the coupling strength.  However, this approximation is
not well justified in $sp$ metals and neglects distinctions between bands of
different character that we expect to be crucial for high T$_c$.
If one assumes the wavevector dependence of coupling is not strong,
there is another simple way to 
identify strong coupling using
deformation potentials ${\cal D} \equiv \Delta 
\varepsilon_k/\Delta Q$ due to frozen-in 
phonon modes with mode amplitude $Q$.\cite{khan}  The underlying
concept is that a phonon that is strongly coupled to Fermi surface states
will produce a large shift in $\varepsilon_k$ for states near the
Fermi level.\cite{khan}

\begin{figure}[t]
\epsfysize=7.0cm\centerline{\epsffile{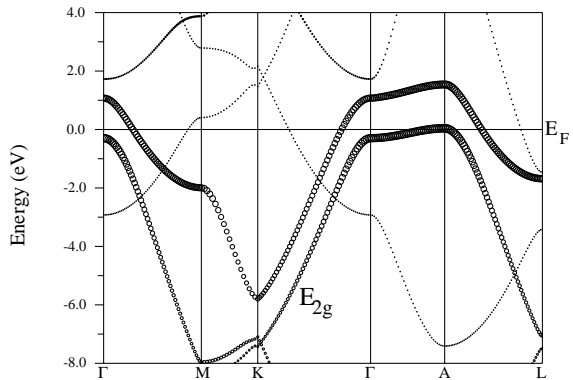}}
\vskip -10mm
\caption{Band structure with frozen-in E$_{2g}$ mode, plotted along
the same lines as in Fig. 1 to facilitate comparison.  (Note that
the point labelled K is no longer a symmetry point.)
The E$_{2g}$ phonon breaks the symmetry and splits the $\sigma$
bands by 1.5 eV along $\Gamma$-A.
}
\label{Bands2}
\end{figure}

We have studied these deformation potentials for the $k$=0 phonons
B$_{1g}$, E$_{2g}$, A$_{2u}$, E$_{1u}$ 
whose energies, 86, 58, 48, 40 meV respectively,
have been calculated by Kortus {\it et al.}\cite{kortus}   
This mode (and others) destroy the symmetry of the
crystal.  The Brillouin zone remains unchanged, however, and we plot the
bands along the same directions as in Fig. \ref{Bands1}.
This mode, as well as the B$_{1g}$ mode, involves
only out-of-phase motion of the two B atoms in the primitive cell, E$_{2g}$
involving in-plane vibrations (bond stretching), B$_{1g}$ having displacements
perpendicular to the plane.  In Fig. \ref{Bands2} the actual rms B 
displacement of $\Delta u_B$ = 0.057 \AA~was used,
to provide a clear picture of the very large effect of
EP coupling strength.   

\vskip -7mm

The E$_{2g}$ phonon strongly splits the $\sigma$ band 
nearly uniformly all along the $\Gamma$-A line
and near $\varepsilon_F$, with the ``gap'' opening
$\Delta \varepsilon_{gap}/\Delta u_B$ =
26 eV/\AA = 2${\cal D}_{E_{2g}}$; 
note that it will be the square of ${\cal D}$ that enters the
expression (below) for $\lambda$.  
In stark contrast, the B$_{1g}$, A$_{2u}$, and E$_{1u}$ modes 
produce no visible effect in
the bands; we estimate that their
deformation potentials are at least a factor of 25 smaller. 
Thus the only
significant deformation potential is for the $\sigma$ bands ($\pi$ band
shifts are always small), and only due to the E$_{2g}$
mode.  This one is extremely large, suggesting that non-linear coupling may
even be occurring.  This strong coupling should be
observable as a large E$_{2g}$ linewidth, and the superconducting gap 
can be expected to be larger on the $\sigma$ Fermi surface sheets than
on the $\pi$ sheets. 

To estimate the coupling from this mode alone, we use 
Eq. (2.34) of Kahn and Allen\cite{khan} for the EP matrix
element $M$ in terms of ${\cal D}_{E_{2g}}$ = 13 eV/\AA,
including only the two $\sigma$ bands N$_{\sigma}$ = 0.11 eV$^{-1}$
per spin (from Fig. 2) and the two E$_{2g}$ modes:  
\begin{eqnarray}
\lambda_{E_{2g}} & = & N_h(\varepsilon_F) \sum_{\nu=1,2}
  \Bigl< \frac{|M_{k,k}^{\nu}|^2}{\omega_{\nu}} \Bigr>_{\varepsilon_F}\\
        &  = & 2
 N_h(\varepsilon_F) \Bigl[\frac{\hbar}{2M_B \omega^2}\Bigr]
 |\sum_{j=1,2}\hat \epsilon_j \cdot \vec {\cal D}_j|^2 \approx 1.0.
\end{eqnarray}
The sum on $j$ runs over the two moving B atoms of the E$_{2g}$ mode.
Using this value of $\lambda$, the E$_{2g}$ mode frequency, 
and the Allen-Dynes\cite{pba} or
McMillan equation with Coulomb pseudopotential $\mu^*$ = 0.10 - 0.15,
the resulting range is T$_c$ = 32-46 K.  There is considerable 
uncertainly in this estimate,
but it seems quite plausible that coupling of the $\sigma$ band
to the bond stretching mode may provide most of the coupling to
account for the observed value
of T$_c$.

Now we summarize, beginning with the effects of low dimensionality 
mentioned in the introduction: (1) the B$_2$ layers
provide a strong differentiation between B states ($\sigma$ vs. $\pi$)
that results in the Mg$^{2+}$ layer giving a 3.5 eV $\sigma - \pi$ energy
shift, driving self-doping of the $\sigma$ bands,
(2) due to their 2D dispersion,
the contributions of the $\sigma$ bands to
N($\varepsilon_F$) is almost independent of the doping level; specifically,
for low doping level N$-h$($\varepsilon_F$) is {\it not} small.  Then, the
strong covalent nature of the $\sigma$ bands leads to an extremely large
deformation potential for the bond stretching modes.  Judging from the 
very small deformation potentials of the other three $\Gamma$ point modes,
the bond-stretching modes dominate the coupling.

W.E.P. is grateful to P. C. Canfield for conversations and for
early communication
of manuscripts.  This work was supported by Office of Naval Research 
Grant N00017-97-1-0956.

{\it Note added}: Since this paper was submitted, two reports of EP coupling
in MgB$_2$ that support our results\cite{EP1,EP2} have come to our
attention.

\vskip -5mm


\begin{references}
\bibitem{gavaler}J. R. Gavaler, Appl. Phys. Lett. {\bf 23}, 480 (1973).
\bibitem{bkbo}L. F. Mattheiss, E. M. Gyorgy, and D. W. Johnson Jr.,
  Phys. Rev. B {\bf 37}, 3745 (1988); R. J. Cava and B. Batlogg,
  MRS Bulletin {\bf 14}, 49 (1989).
\bibitem{cs3c60}T. T. M. Palstra {\it et al.}, Solid State Commun.
  {\bf 93}, 327 (1995).
\bibitem{MNCl}S. Yamanaka {\it et al.},
  Adv. Mater. {\bf 9}, 771 (1996); S. Yamanaka, K. Hotehama and H. Kawaji, 
  Nature {\bf 392}, 580 (1998);
  S. Shamoto {\it et al.},
  Physica C {\bf 306}, 7 (1998).
\bibitem{c60}J. H. Schon, C. Kloc, and B. Batlogg, Nature {\bf 408},
  549 (2000).
\bibitem{akimitsu}J. Akimitsu, {\it Symp. on Transition Metal Oxides},
  Sendai, January 10, 2001: J. Nagamatsu {\it et al.} Nature {\bf 410},
  63 (2001).
\bibitem{budko}S. L. Bud'ko {\it et al.}, Phys. Rev. Lett. {\bf 86},
  1877 (2001).
\bibitem{other}
  D. K. Finnemore {\it et al.}, cond-mat/0102114;
  G. Rubio-Bollinger {\it et al.}, cond-mat/0102242;
  B. Lorenz {\it et al.}, cond-mat/0102264;
  P. C. Canfield {\it et al.}, cond-mat/0102289;
  A. Sharoni {\it et al.}, cond-mat/0102325;
  H. Kotegawa {\it al}, cond-mat/0102334.
\bibitem{c2}I. T. Belash {\it et al.}, Solid State Commun. {\bf 64},
  1445 (1987).
\bibitem{kortus}J. Kortus {\it et al.}, cond-mat/0101446.
\bibitem{satta} G. Satta {\it et al.}, cond-mat/0102358.
\bibitem{cava}J. S. Slusky {\it et al.}, cond-mat/0102262.
\bibitem{hirsch}J. E. Hirsch, cond-mat/0102115.
\bibitem{djsbook}D. J. Singh, {\it Planewaves, Pseudopotentials, and the
LAPW Method} (Kluwer Academic, Boston, 1994).
\bibitem{wien}P. Blaha, K. Schwarz, and J. Luitz, WIEN97, Vienna
University of Technology, 1997. Improved and updated version of the
original copyrighted WIEN code, which was published by P. Blaha,
K. Schwarz, P. Sorantin, and S. B. Trickey, Comput. Phys. Commun.
{\bf 59}, 399 (1990).
\bibitem{gga}J. P. Perdew {\it et al.}, Phys. Rev. B {\bf 46}, 6671 (1992);
J. P. Perdew, K. Burke, and M. Ernzerhof, Phys. Rev. Lett. {\bf 77},
3865 (1996).
\bibitem{bela}K. D. Belashchenko, M. van Schilfgaarde, and V. P.
  Antropov, cond-mat/0102290.
\bibitem{GG}G. D. Gaspari and B. L. Gyorffy, Phys. Rev. Lett. {\bf 28},
  801 (1972).
\bibitem{khan}F. S. Khan and P. B. Allen, Phys. Rev. B {\bf 29}, 3341 (1984).
\bibitem{pba}P. B. Allen and R. C. Dynes, Phys. Rev. B {\bf 12}, 905 (1975).
\bibitem{EP1}A. Y. Liu, J. Kortus, and I. I. Maxin, unpublished.
\bibitem{EP2}Y. Kong {\it et al.}, cond-mat/0102499.
\end{references}
\end{document}